\documentclass[12pt]{article}

\usepackage{amsmath}
\usepackage{amssymb}
\usepackage{amsthm}
\usepackage{bm}

\theoremstyle{definition}
\newtheorem{theorem}{Theorem}
\newtheorem{lemma}{Lemma}

\title{
Extension of the Lieb-Schultz-Mattis and Kolb theorem
}
\author{Kiyohide Nomura
\\
Department of Physics, Kyushu University
\\
Fukuoka 819-0395,
JAPAN
}
\begin{document}
\maketitle

\begin{abstract}
 
The theorem of Lieb, Schultz and Mattis (LSM)
 \cite{Lieb-Schultz-Mattis-1961},
 which states that
 the S=1/2 XXZ spin chain has gapless or degenerate ground states,
can be applied to broader models.
Independently, Kolb \cite{Kolb-1985} considered
the relation between the wave number $q$ and the twisting boundary
 condition,
 and he obtained a similar result as LSM. 
 However, in frustrating cases it is known that
 there exist several exceptions
 for the assumption of the {\em unique lowest
 state for the finite size},
 which is important in the traditional LSM theorem.
 In our previous paper,
without the assumption of the uniqueness,
 we have extended the LSMK theorem for frustrating and
 non-symmetric cases.
However, there remains a complexity in the proof of continuity. 
 In this paper, we will simplify the proof than the previous work.

\end{abstract}

\section{Introduction}

In statistical physics, rigorous theorems play important role;
the Mermin-Wagner theorem, the Marshall-Lieb-Mattis theorem
\cite{Marshall-1955,Lieb-Mattis-1962},
 the Lieb-Schultz-Mattis theorem \cite{Lieb-Schultz-Mattis-1961}
 etc,
 which do not give quantitative but qualitative results,
and can be applied to broad models.
 And one can use them to check the consistency of approximations,
 experiments, or numerical data.

 Lieb, Schultz and Mattis (LSM)
\cite{Lieb-Schultz-Mattis-1961}
studied the S=1/2 XXZ spin chain.
In appendix B of \cite{Lieb-Schultz-Mattis-1961}, they stated two theorems.
For the finite $L$ size, 
the uniqueness of the ground state was proved in the first theorem.
In the second theorem, they proved
that there exists a low-energy $O(1/L)$
 excited state;
 in the infinite limit,
this means that either there are degenerate ground states or a vanishing gap.
The first theorem was nothing more than an extension of Marshall's theorem
\cite{Marshall-1955}, later generalized \cite{Lieb-Mattis-1962},
therefore it is appropriate to call the first theorem
the ``Marshal-Lieb-Mattis (MLM) theorem''.
However the MLM theorem 
is limited in non-frustrated cases.
The second LSM theorem was extended for general spin $S$ 
and was applied for various models \cite{Affleck-Lieb-1986,Affleck-1988}.

For rational magnetizations, 
by using the LSM discussion,
Oshikawa {\it et al.}
\cite{Oshikawa-Yamanaka-Affleck-1997} 
pointed out that there are multiple degenerate energy states
with a gap in the infinite limit
(relating to the spontaneous translation symmetry breaking)
or gapless. 
In addition, they emphasized 
the discrete symmetry (the space inversion {\em or} the spin
reversal {\em or} the time reversal symmetry),
besides the U(1) and the translation symmetry.
Although the discrete symmetry simplifies the proof of the LSM theorem,
it excludes non-symmetric cases. 

Independently of the LSM discussions
\cite{Lieb-Schultz-Mattis-1961,Affleck-Lieb-1986},
Kolb \cite{Kolb-1985} studied the energy spectra of the XXZ spin
chain with the twist boundary condition.
He pointed out the shift of the (pseudo) wave number $q$ when varying the twist
boundary condition. 
For the $S$ half-odd-integer case.
he showed the nontrivial periodicity ($q\rightarrow q+\pi$)
of the energy spectra in the Brillouin zone,
which means the two-fold pseudo degenerate ground
states. 
In section II of \cite{Fath-Solyom-1993},
F\'ath and S\'olyom combined the Kolb's idea and LSM theorem,
and they argued the continuity of the energy spectra
for $S^{z}_{T} = \pm 1,\pm 2, \cdots$.

One limitation of the traditional LSM theorem
is the assumption of the {\em unique ground state for the finite size}
(or the {\em unique lowest energy state} 
in the fixed magnetization subspace).  
However, when including frustrations,
there exist counterexamples for this assumption;
one is the double-fold ground states in the Majumdar-Ghosh model
\cite{Majumdar-Ghosh-1969}
($\alpha=1/2$ in (\ref{eq:S=1/2-NNN-XXZ-chain}))
\begin{align}
 \hat{H} =  \sum_{j}  \hat{\bm{S}}_{j}\cdot\hat{\bm{S}}_{j+1}
 + \alpha  \hat{\bm{S}}_{j}\cdot\hat{\bm{S}}_{j+2}
\label{eq:S=1/2-NNN-XXZ-chain}
\end{align}
(more generally multi-fold lowest states in matrix product models),
another is the double-well energy spectrum
observed in one spin flip from the fully aligned
state in the incommensurate region
($\alpha>1/4$
in (\ref{eq:S=1/2-NNN-XXZ-chain})).

In our previous paper
\cite{Nomura-Morishige-Isoyama-2015},
we separated the LSM theorem from the MLM one,
without the assumption of the {\em uniqueness} of the ground state,
using a squeeze theorem type method,
and combing the LSM theorem with Kolb's discussion
(hereafter we call the LSM theorem with Kolb's one as LSMK theorem).
Also we tried not to use the {\em discrete} symmetry in the proof process.
Therefore we can extend the LSMK theorem for
frustrating or non-symmetric models. 
However, in our previous work,
the proof of the continuity of energy spectra
was not straightforward.
In this paper, we will polish it.

The layout of the paper as follows.
In section 2, we introduce the definition of symmetry operations.
Section 3 is the main part of this work:
we prove the continuity and the nontrivial periodicity of the lowest energy spectra 
as a function of wave number $q$,
assuming the U(1) and the translational symmetries
plus the short-range interaction.
Section 4 is the conclusion.

\section{Model, symmetries, eigenstates}

In this section we consider the symmetries of the spin chain.
As a typical model, we treat the following 
generalized XXZ spin chain:
\begin{align}
 \hat{H} &= \sum_{j=1}^{L} \sum_{r=1}^{HL}
 \left(
 J(r)(\hat{S}^{x}_{j}  \hat{S}^{x}_{j+r}+\hat{S}^{y}_{j} \hat{S}^{y}_{j+r} )
  + \Delta(r)  \hat{S}^{z}_{j} \hat{S}^{z}_{j+r}
\right)+ h\sum_{j=1}^{L} \hat{S}^{z}_{j}
\notag \\
  &= \sum_{j=1}^{L} \sum_{r=1}^{HL}
\left( \frac{J(r)}{2}(\hat{S}^{+}_{j}
  \hat{S}^{-}_{j+r}+\hat{S}^{-}_{j} \hat{S}^{+}_{j+r}  )
  + \Delta(r)  \hat{S}^{z}_{j} \hat{S}^{z}_{j+r}
\right)+ h\sum_{j=1}^{L} \hat{S}^{z}_{j},
\label{eq:Hamiltonian-generalized-XXZ}
\end{align}
where
$
 (\hat{\bm{S}}_{j})^{2}= S(S+1)
$
($S=1/2,1,\cdots$),
$L$ denotes the system size, $HL=[L/2]-1$,
and the periodic boundary condition (PBC).
We can also include multibody 
 and nonsymmetric interactions, as will be shown later.
      
\subsection{Symmetries}

Next we enumerate the symmetry operations.
Hereafter we denote
\begin{equation}
\hat{S}^{x}_{T} \equiv \sum_{j=1}^{L} \hat{S}^{x}_{j},
\quad
\hat{S}^{y}_{T} \equiv \sum_{j=1}^{L} \hat{S}^{y}_{j},
\quad
\hat{S}^{z}_{T} \equiv \sum_{j=1}^{L} \hat{S}^{z}_{j}.
\end{equation}

\begin{enumerate}
 \item{Rotation operator around the $z$-axis:
      $\hat{U}^{z}_{\theta} \equiv  \exp(- i \theta   \hat{S}^{z}_{T})$.}

      The rotation operator satisfies:
      \begin{align}
  (\hat{U}^{z}_{\theta})^{\dagger} \hat{S}^{\pm}_{j} \hat{U}^{z}_{\theta}
  =
\hat{S}^{\pm}_{j} \exp(\pm i\theta),
\quad
(\hat{U}^{z}_{\theta})^{\dagger} \hat{S}^{z}_{j} \hat{U}^{z}_{\theta} =  \hat{S}^{z}_{j}.
      \end{align}

\item{Translation operator by one-site: $\hat{U}_{\rm trl}$.} 

\begin{equation}
 \hat{U}_{\rm trl}^{\dagger} \hat{S}^{x,y,z}_{j}  \hat{U}_{\rm trl} = 
 \hat{S}^{x,y,z}_{j+1}.
\end{equation}

 \item

The operators 
$\hat{S}^{x,y,z}_{T} $
are invariant under the translation. 

\end{enumerate}

\subsection{Eigenstates}
We write the eigenstate for the total spin $\hat{S}^{z}_{T}$ and the
translation:
\begin{align}
\hat{S}^{z}_{T}|S^{z}_{T};  q\rangle = S^{z}_{T} |S^{z}_{T};  q\rangle,
\quad
 \hat{U}_{\rm trl}|S^{z}_{T}; q \rangle = \exp(iq) |S^{z}_{T};  q\rangle,
\end{align}
where the total spin eigenvalue is related with the magnetization as
$M \equiv S^{z}_{T}/L$, and $q$ is the wave number.

Since the Hamiltonian is U(1) and translational invariant,
one can choose
\begin{align}
 \hat{H} |S^{z}_{T};  q\rangle &= E(S^{z}_{T};  q) |S^{z}_{T};  q\rangle.
\end{align}
Energy spectra are $2\pi$ periodic with the wave number $q$:
\begin{equation}
 E(S^{z}_{T};  q+2\pi) = E(S^{z}_{T};  q).
  \label{eq:Energy-spcetra-periodicity}
\end{equation}

\section{Extension of the LSMK theorem}

In this section, we will extend the LSMK theorem
without the assumption of the uniqueness of the lowest energy,
by using squeeze theorem type methods.
And we will use only the U(1) and the translational
symmetry.  
We do not assume the discrete symmetry such as
the space inversion or the spin reversal or the time reversal.
Hereafter we express $|S^{z}_{T}; q\rangle$
as one of the lowest energy eigenstates in the subspace of $S^{z}_{T}$
and $q$ for the Hamiltonian $\hat{H}$ with PBC.

In the subsection 3.1,
we define the twisted boundary condition (TBC) and the twisting operator;
although in this paper we will only use PBC,
for the purpose of the Taylor expansion (\ref{eq:Taylor-expansion})
later, we mention TBC.
In the subsection 3.2, we review the LSMK theorem
according to
\cite{Nomura-Morishige-Isoyama-2015,Barwinkel-Hage-Schmidt-Schnack-2003,Hakobyan-2003},
since the formalism in \cite{Kolb-1985, Fath-Solyom-1993} was somewhat
cumbersome because of the wave-function treatment,
and higher order calculations become simpler than
\cite{Affleck-Lieb-1986}
when multibody interactions are included.
Using the squeezing method,
we discuss the periodicity for the rational magnetization in 3.3,
and the continuity for the irrational magnetization in 3.4.

\subsection{Twisted boundary condition and twisting operator}

We introduce the
twisted boundary condition (TBC):
\begin{equation}
 \hat{S}^{\pm}_{L+j} =\hat{S}^{\pm}_{j}\exp(\pm i\Phi), \;
 \hat{S}^{z}_{L+j} =\hat{S}^{z}_{j},
\end{equation}
and we shall denote the Hamiltonian
(\ref{eq:Hamiltonian-generalized-XXZ})
with TBC as $\hat{H}_{\Phi}$.

Next we define the twisting unitary operator
\cite{Nomura-Morishige-Isoyama-2015,Barwinkel-Hage-Schmidt-Schnack-2003}
:
\begin{equation}
 \hat{U}^{\rm tw}_{\Phi} \equiv \exp \left(-i \frac{\Phi}{L} \sum_{j=1}^{L} j (\hat{S}^{z}_{j}
				      -S)\right),
\label{eq:Twisting-Operator}
\end{equation}
then we obtain
\begin{align}
(\hat{U}^{\rm tw}_{\Phi})^{\dagger} \hat{S}^{\pm}_j \hat{U}^{\rm tw}_{\Phi}
=  \hat{S}^{\pm}_j \exp( \pm i\Phi j /L),
\;
(\hat{U}^{\rm tw}_{\Phi})^{\dagger} \hat{S}^{z}_j \hat{U}^{\rm tw}_{\Phi}
=  \hat{S}^{z}_j, 
\end{align}
and
$ [\hat{U}^{\rm tw}_{\Phi},\hat{U}^{z}_{\theta}]=0$.

Applying the twisting operator (\ref{eq:Twisting-Operator})
for $\hat{H}_{\Phi}$, 
we obtain
\begin{align}
 (\hat{U}^{\rm tw}_{\Phi})^{\dagger} \hat{H}_{\Phi}  \hat{U}^{\rm tw}_{\Phi}
 -\hat{H}_{\Phi}
 = \sum_{j=1}^{L} \sum_{r=1}^{HL} 
 \frac{J(r)}{2}
 (\hat{S}^{+}_{j} \hat{S}^{-}_{j+r} (\exp(- i \frac{\Phi r}{L})-1)
+\text{h. c.}).
 \label{eq:Hamiltonian-TBC}
\end{align}
Note that we will only use this expression 
in the Taylor expansion (\ref{eq:Taylor-expansion}) around $\Phi=0$
to $\Phi=2\pi l$($l$: integer) (PBC).

\subsection{LSM theorem and translation operator}

\begin{lemma}
 (Translation operator and twisting operator)

\begin{align}
\hat{U}^{\rm tw}_{2\pi l} \hat{U}_{\rm trl}
&=  \hat{U}_{\rm trl}\hat{U}^{\rm tw}_{2\pi l}
 \exp\left(\frac{2 \pi i l}{L}(\hat{S}^{z}_{T}
 -SL)\right).
 \quad (l: \text{integer})
 \label{eq:TBC-translation-2pi}
\end{align}

\begin{proof}
 
\begin{align}
\hat{U}_{\rm trl}^{\dagger} \hat{U}^{\rm tw}_{\Phi} \hat{U}_{\rm trl}
&=\exp\left(- \frac{i \Phi}{L}\sum_{j=1}^{L} j(\hat{S}^{z}_{j+1} -S)\right)
\notag \\
&=\exp\left(-\frac{i \Phi}{L}\left(\sum_{j=2}^{L} (j-1)(\hat{S}^{z}_{j} -S)
+L(\hat{S}^{z}_{L+1}-S)\right)
\right)
\notag \\
&=\hat{U}^{\rm tw}_{\Phi} \exp\left(\frac{i \Phi}{L}(\hat{S}^{z}_{T}
 -SL)\right)\exp(- i \Phi (\hat{S}^{z}_{1} -S)),
 \label{eq:TBC-translation-2pi-2}
\end{align}
 where we used  $\hat{S}^z_{L+1} = \hat{S}^z_{1} $.
By setting $\Phi=2\pi l$ and using the fact that
the eigenvalue of $\hat{S}^{z}_{1} -S$ is an integer,
we obtain
(\ref{eq:TBC-translation-2pi}).

\end{proof}
\end{lemma}

\begin{theorem}

In the subspace with a quantum number $S^{z}_{T}$,
on the lowest energies of the two wave numbers 
$q$ and $q-2\pi l S^{z}_{T}/L +2\pi l S$
($l$ is an integer $|l| \ll L$),
 the next inequality holds:
\begin{equation}
E(S^{z}_{T};  q-2\pi l S^{z}_{T}/L +2\pi l S)
- E(S^{z}_{T}; q)
\le O(l^{2}/L).
\label{eq:Energy-Spectrum-Inequality}
\end{equation}

\begin{proof}
 
 The following combination
\begin{align}
(\hat{U}^{\rm tw}_{2\pi l})^{\dagger} \hat{H} \hat{U}^{\rm tw}_{2\pi l}  
 -\hat{H},
 \label{eq:Double-Twisting-Hamiltonian}
\end{align}
is translational invariant from the Lemma 1.
Also from the Lemma 1, we obtain
 \begin{equation}
  \hat{U}_{\rm trl} (\hat{U}^{\rm tw}_{2\pi l}|S^{z}_{T}; q\rangle)
   = \exp(i(q - 2\pi l S^{z}_{T} /L + 2\pi l S))
   (\hat{U}^{\rm tw}_{ 2\pi l}|S^{z}_{T}; q\rangle).
   \label{eq:Wave-number-eigenstate-of-twisting}
 \end{equation}
%since $2S$ is an integer and (\ref{eq:Energy-spcetra-periodicity}).

 Next we consider the Taylor expansion:
\begin{align}
 (\hat{U}^{\rm tw}_{\Phi})^{\dagger} \hat{H}_{\Phi}
   \hat{U}^{\rm tw}_{\Phi} 
   =  \hat{H}_{\Phi=0}
+ \Phi 
   \left[\frac{d }{d \Phi}\left( (\hat{U}^{\rm tw}_{\Phi})^{\dagger} \hat{H}_{\Phi}
  \hat{U}^{\rm tw}_{\Phi} \right)\right]_{\Phi =0}
 + O(\Phi^{2}).
 \label{eq:Taylor-expansion}
\end{align}
 By the way, using the next relation 
 \begin{align}
\left[\frac{d }{d \Phi}\left( (\hat{U}^{\rm tw}_{\Phi})^{\dagger} \hat{H}_{\Phi}
  \hat{U}^{\rm tw}_{\Phi} \right)\right]_{\Phi =0}
  = - \left[\hat{H},\frac{ i }{L}\sum_{j=1}^{L}j(\hat{S}^{z}_{j}-S) \right],
 \end{align}
and the fact
\begin{align}
 \langle S^{z}_{T}; q|
 [\hat{H},\sum_{j=1}^{L}j(\hat{S}^{z}_{j}-S)]
 | S^{z}_{T}; q \rangle
 =0,
\end{align}
 we obtain
 \begin{align}
  \langle S^{z}_{T}; q|
  \left[
  \frac{d }{d \Phi}\left( (\hat{U}^{\rm tw}_{\Phi})^{\dagger} \hat{H}_{\Phi}
  \hat{U}^{\rm tw}_{\Phi} \right)\right]_{\Phi =0}
   | S^{z}_{T}; q \rangle
 =0.
\label{eq:Expectation-of-1st-twisting-term}
 \end{align}

Using equations (\ref{eq:Double-Twisting-Hamiltonian}),
 (\ref{eq:Wave-number-eigenstate-of-twisting}),
  (\ref{eq:Taylor-expansion}) and
 (\ref{eq:Expectation-of-1st-twisting-term}) for models (\ref{eq:Hamiltonian-generalized-XXZ}),
we can prove the following inequality:
\begin{align}
& E(S^{z}_{T}; q - 2\pi l S^{z}_{T} /L + 2\pi l S)
-  E(S^{z}_{T}; q)
\notag \\
& \le 
\langle  S^{z}_{T}; q  | 
((\hat{U}^{\rm tw}_{2\pi l})^{\dagger} \hat{H} \hat{U}^{\rm tw}_{2\pi l}  
- \hat{H} )
| S^{z}_{T}; q  \rangle
\notag \\
&
=
\sum_{j=1}^{L}\sum_{r=1}^{HL} J(r)
\kappa\left(\frac{ r l}{L}\right)
 \langle  S^{z}_{T}; q  |
\hat{S}^{+}_{j}\hat{S}^{-}_{j+r}
|S^{z}_{T}; q  \rangle + \text{h. c.}
\notag \\&
\le
2 \sum_{j=1}^{L}\sum_{r=1}^{HL} |J(r)|
\left| \kappa\left(\frac{ r l}{L}\right) \right|
 |\langle  S^{z}_{T}; q  |
\hat{S}^{+}_{j}\hat{S}^{-}_{j+r}
|S^{z}_{T}; q  \rangle| 
\le O(l^{2}/L),
\label{eq:Energy-Spectrum-Inequality2}
\end{align}
 where
 $
 \kappa(\phi)\equiv \frac{1}{2} (\exp (-2\pi i \phi) -(1-2\pi i\phi))
 \approx O(\phi^2)
 $\cite{Hakobyan-2003}.
 
In the course of proof, we have used the variational principle,
the translational invariance,
the boundedness of spin operators 
 $
|\langle  \hat{S}^{+}_{j}\hat{S}^{-}_{j+r} \rangle|
 \le 4S^{2}
 $,
  and that the transverse interaction is short-range
 (for details, see appendix).
\end{proof}
\end{theorem}

[Remarks]
\begin{enumerate}
\item
Although the form of
(\ref{eq:Energy-Spectrum-Inequality2})
seems specific for the model (\ref{eq:Hamiltonian-generalized-XXZ}),
one can prove it for general U(1) symmetric models;
the multibody interactions:
$
      (\hat{\bm{S}}_{j} \cdot \hat{\bm{S}}_{j+r_{1}})
             (\hat{\bm{S}}_{j+r_{2}} \cdot \hat{\bm{S}}_{j+r_{3}})
$
and the Dzyaloshinskii-Moriya type interaction:
$
 (\hat{\bm{S}}_{j}\times \hat{\bm{S}}_{j+1})_{z}
$
     etc.
They are expressed as a sum of terms
\begin{equation}
  \hat{S}^{+}_{j}  \hat{S}^{-}_{j+r_1}
   \hat{S}^{+}_{j+r_2}  \hat{S}^{-}_{j+r_3}\cdots,
\end{equation}
where the number of the raising operators must be equal to the number of
the lowering operators from the U(1) symmetry.
Then it is easy to show the inequality 
(\ref{eq:Energy-Spectrum-Inequality2}).

 \item 
The longitudinal interaction $\Delta(r)$ and the magnetic field $h$
give no restriction on Theorem 1.

 \item
      One can prove a similar result as the Lemma 1 for the fermion  \cite{Yamanaka-Oshikawa-Affleck-1997,Gagliardini-Hass-Rice-1998}
      and the boson \cite{Oshikawa-2000}.
      However, in the proof process of the Theorem 1,
      we have used the boundedness of operators.
      Thus, it is safe to apply the LSM theorem for interacting
      fermions systems on a lattice
      \cite{Yamanaka-Oshikawa-Affleck-1997,Gagliardini-Hass-Rice-1998},
      whereas for the boson operator which is not bounded,
      one can not prove the LSM theorem.
\end{enumerate}

\subsection{Nontrivial periodicity of energy spectra for rational magnetizations}

\begin{theorem}

The lowest energy spectrum in the subspace
of $S^{z}_{T}=(S-m/n)L$ ($m,n$ are coprime integers, independent of
 $L$)
is non-trivially periodic as $q\rightarrow q+2\pi/n$
 in the infinite limit:
 \begin{equation}
  \lim_{L\rightarrow\infty}|E(S^{z}_{T};  q) - E(S^{z}_{T};  q+2\pi/n) | =0.
 \end{equation}
 
\begin{proof}
 
From the Theorem 1, we obtain
\begin{align}
& E(S^{z}_{T};  q + 2\pi m l/n )
-  E(S^{z}_{T};  q)
\le O(l^{2}/L),
\end{align}

 Secondly, applying the Theorem 1 to the lowest energy state
 with $q + 2\pi m l/n$, we obtain      
\begin{align}
 E(S^{z}_{T};  q)
-  E(S^{z}_{T};  q + 2\pi m l/n )
\le O(l^{2}/L),
\end{align}
therefore
\begin{equation}
  |E(S^{z}_{T};  q) - E(S^{z}_{T};  q +  2\pi m l/n) | \le O(l^{2}/L).
 \end{equation}    

Finally, since $m,n$ are coprime, one can choose integers $l,k$:
\begin{equation}
 m l + n k= 1,
\end{equation}
and remembering (\ref{eq:Energy-spcetra-periodicity}), we obtain
 \begin{equation}
  \left| E(S^{z}_{T};  q) - E(S^{z}_{T};  q +  2\pi/n) \right|
   \le O(l^{2}/L).
 \end{equation}    
  \end{proof}
\end{theorem}

\medskip

[Remarks]
\begin{enumerate}
\item Naively, the number of minima of the lowest energy
      spectrum
      in the Brillouin zone ($-\pi \le q <\pi$)  
      should be $n$ from the Theorem 2.
      However, there exist the cases where the number of minima is $2n, 3n,\cdots$.
 \item 
The nontrivial periodicity of the Theorem 2 is valid only for the lowest energy spectrum.
\end{enumerate}

\subsection{Continuity of energy spectra for irrational magnetizations}

  \begin{theorem}
   
The lowest energy spectrum in the subspace
of $S^{z}_{T}=(S-m/n)L+\Delta S^{z}_{T}$ \;
($m, n$ are coprime integers, independent of
 $L$; \; $\Delta S^{z}_{T}$ is an integer with $|\Delta S^{z}_{T}| \ll L$)
  is continuous as a function of the wavenumber $q$ 
   in the infinite limit,
   except $\Delta S^{z}_{T}=0$.

\begin{proof}

By taking the twist operator as
 $\hat{U}^{\rm tw}_{2\pi n}$,
 and using the Theorem 1
with (\ref{eq:Energy-spcetra-periodicity}),
we obtain
 \begin{equation}
  E(q+ 2\pi n \Delta S^{z}_{T}/L ) -  E(q) \le O(n^{2}/L).
 \end{equation}
 Conversely we can show
 \begin{equation}
 E(q)-  E(q+ 2\pi n\Delta S^{z}_{T}/L ) \le O(n^{2}/L).
 \end{equation}
 Therefore we obtain
 \begin{equation}
  \left| E(q)-  E(q+ 2\pi n\Delta S^{z}_{T}/L ) \right|
   \le O(n^{2}/L),
 \end{equation}
 that is, the lowest energy spectrum is continuous
 in the infinite limit. 
 
\end{proof}
  \end{theorem}

[Remarks]
\begin{enumerate}
 \item 
One cannot prove the continuity of the lowest energy spectrum in
the $S^{z}_{T}=(S-m/n)L$ subspace.

\item
Although the lowest energy spectrum of
$\Delta S^{z}_T =\pm 1,\pm 2, \cdots$
is continuous,
the derivative of the spectrum may be discontinuous.
\end{enumerate}

 \section{Conclusion}

We have extended the LSMK theorem including the frustrated case,
because we have not used the uniqueness condition of the lowest state in
each $S^{z}_{T}$ subspace.
We have also extended the LSMK theorem for the non-symmetric case, 
for example, the Dzyaloshinskii-Moriya interaction.

Although there are many researches on the traditional LSM theorem,
almost all of them have assumed the {\em uniqueness of the lowest energy
state for the finite system}
or more restrictively the MLM (Perron-Frobenius) theorem  
\cite{Lieb-Schultz-Mattis-1961,Affleck-Lieb-1986,Affleck-1988,Oshikawa-Yamanaka-Affleck-1997,Barwinkel-Hage-Schmidt-Schnack-2003,Hakobyan-2003,Rojo-1996,Fukui-Sigrist-Kawakami-1997,Yamanaka-Oshikawa-Affleck-1997,Gagliardini-Hass-Rice-1998}.
F\'ath and S\'olyom \cite{Fath-Solyom-1993} did not mention explicitly
the assumption of the {\em uniqueness} of the lowest energy state in the fixed
magnetization,
however, they did not discuss carefully
to avoid the uniqueness assumption.
Although in the original statement
of LSM \cite{Lieb-Schultz-Mattis-1961,Affleck-Lieb-1986},
the unique lowest state assumption is harmless,
for further applications of the LSM theorem,
for example, the magnetic plateaux \cite{Oshikawa-Yamanaka-Affleck-1997}
where a multi-fold degeneracy may occur, 
or the continuity of the energy spectra
\cite{Fath-Solyom-1993},
this assumption may become an obstacle.
Interestingly, 
Affleck et al. \cite{Affleck-Lieb-1986,Affleck-1988}
touched that the assumption of a unique ground state can
fail for some cases, but did not discuss profoundly.

Another assumption, the {\em discrete symmetry} such as
the space inversion {\em or} the spin reversal {\em or} the time reversal,
was introduced in
\cite{Oshikawa-Yamanaka-Affleck-1997}
and has been widely used
\cite{Fukui-Sigrist-Kawakami-1997,Yamanaka-Oshikawa-Affleck-1997,Gagliardini-Hass-Rice-1998}.
Although these discrete symmetries simplify
the proof of the LSMK theorem (especially for multibody interactions),
they exclude the non-symmetric interactions.
However, reexamining the proof process,
these discrete symmetries are not needed
in the LSMK theorem (the first one who noticed this point was
\cite{Koma-2000}).

Another by-product of the separation of the MLM theorem and the discrete
symmetry from the LSMK theorem is that the requirement
of the evenness of system size $L$ can be omitted.

When the magnetization is irrational,
	      i.e.,
      $S^{z}_{T}=(S-m/n)L+\Delta S^{z}_{T} \;
(\Delta S^{z}_{T} = \pm 1, \pm 2,\cdots)$,
the lowest energy spectrum 
is continuous for the wave number $q$ 
in the infinite limit.
For the rational magnetization $S^{z}_{T}=(S-m/n)L$,
the lowest energy spectrum has the periodicity
$q\rightarrow q+2\pi/n$ in the wave number space
\cite{Oshikawa-Yamanaka-Affleck-1997},
and the energy spectrum is gapless {\em or} gapped with $n$ (maybe
$2n,3n,\cdots$)-fold degeneracy
(indeed there are $2n$-fold degeneracy cases:
N\'eel state in S=1 XXZ chain etc.).
Note that for the gapless case, there are  $n$ (maybe
$2n,3n,\cdots$) soft modes for the rational magnetization,
from the periodicity $q\rightarrow q+2\pi/n$.

Finally, the original LSM theorem has been applied for
the spin ladder model
\cite{Rojo-1996,Fukui-Sigrist-Kawakami-1997},
the fermion system on the lattice 
\cite{Yamanaka-Oshikawa-Affleck-1997,Gagliardini-Hass-Rice-1998},
and the quantum Hall effect \cite{Koma-2000}.
It will be interesting to consider our methods for fermion
models with frustrations. 
For the interacting boson, although the LSM-like results were expected
\cite{Oshikawa-2000},
one can not prove the LSM theorem because the boson operator is not
bounded;
it would be needed some conditions, for example the hard core (repulsive) interaction,
 to prove the LSM-like results for the interacting boson,
 or boson systems might be intrinsically different from the bounded
 operator cases (spin or fermion).

\section*{Acknowledgment}

We wish to thank T. Koma at Gakushuin University for
critical discussions on the proof of the continuity in \cite{Nomura-Morishige-Isoyama-2015}.

\appendix

\section{Condition on the interaction range}
Here we discuss the condition of the interaction range.
From the latter part of (\ref{eq:Energy-Spectrum-Inequality2}),
we obtain
\begin{align}
        \sum_{r=1}^{HL} |J(r)|
       \left| \kappa\left( r l/L\right) \right|
       =\frac{l^{2}}{L^{2}}
       \sum_{r=1}^{HL} r^{2} |J(r)|
       \frac{\left| \kappa( r l/L)\right|}{(r l/L)^{2} }
       \le  C \frac{l^{2}}{L^{2}}
       \sum_{r=1}^{HL} r^{2} |J(r)|,
\end{align}
where we have used the boundedness
$ | \kappa( \phi)|/\phi^{2} \le C, \; (C>0) $
in the interval $0\le \phi \le l$,
since
$ | \kappa( \phi)|/\phi^{2} $  
is a continuous function.
Therefore, when
\begin{equation}
\lim_{L \rightarrow \infty} 
 \sum_{r=1}^{HL} r^{2} |J(r)| = Constant,
\end{equation}
we obtain the result of Theorem 1.
Moreover, with the weaker condition
\begin{equation}
 \lim_{L \rightarrow \infty} \frac{1}{L}
  \sum_{r=1}^{HL} r^{2} |J(r)| =0,
  \label{eq:Short-range-condition-LSMK}
\end{equation}
we can obtain similar results as Theorems 1,2,3 with slight modifications
(the condition  (\ref{eq:Short-range-condition-LSMK}) is somewhat different from \cite{Hakobyan-2003} ).
Note that (\ref{eq:Short-range-condition-LSMK}) gives only the sufficient
condition for Theorem 1,2,3.
Other model may be gapless or may have degenerate ground states.
The S=1/2 Heisenberg chain with long-range interactions $J(r) =1/r^{2}$
\cite{Haldane-1988, Shastry-1988}
 is such an example.

\end{document}